\documentclass[aps,prd,preprint, nofootinbib,superscriptaddress]{revtex4-1}
\usepackage{array}
\usepackage[table]{xcolor}
\usepackage{soul}
\usepackage[normalem]{ulem}
\usepackage[T1]{fontenc}
\usepackage{comment}
\usepackage{amsmath}
\usepackage{amsthm}
\usepackage{amssymb}
\usepackage{cancel}
\usepackage{graphicx}
\usepackage[unicode=true, bookmarks=false, breaklinks=false, pdfborder={0 0 0},colorlinks=false]{hyperref}
\usepackage{fancyhdr}
\usepackage{epsfig}
\usepackage{slashed}
\usepackage{mathrsfs}
\usepackage[thinc]{esdiff}
\usepackage[titletoc]{appendix}
\usepackage{tikzsymbols}
\usepackage{feynmf}
\usepackage{tikzsymbols}
\usepackage{dsfont}
\usepackage{xcolor}
\usepackage{slashed}
\usepackage{braket}
\usepackage{mathrsfs}
\usepackage{xcolor}
\usepackage{latexsym}
\usepackage[compat=1.1.0]{tikz-feynman}
\usepackage{titlesec}

\titlespacing{\subsection}{0pt}{4pt}{4pt}

\begin{document}
\title{Study of sub-GeV Dipolar Dark States at SND@LHC within Invisible Bounds on Meson Decays}

\author{Debajyoti Biswas}
\email{debajyotib@iisc.ac.in}
\affiliation{Centre for High Energy Physics, Indian Institute of Science, CV Raman Road, Bengaluru 560 012, India}

\date{\today}

\begin{abstract} 
Electromagnetic form factors constitute a natural portal for accessing states beyond the Standard Model. In particular, dimension-5 magnetic and electric dipole moment operators offer a minimal and predictive framework for Feebly Interacting Particles (FIPs). In this work, we study the sensitivity of the Scattering and Neutrino Detector (SND@LHC) to dipolar dark states through photon-mediated interactions with Standard Model particles. The far-forward region of the LHC provides FIPs with large momenta that scatter off electrons and nuclei inside the target. The production of dark states from meson decays is constrained by invisible decay widths, while the Drell-Yan process offers a production channel in the GeV range. We derive projected sensitivity limits for magnetic and electric dipole moment interactions at SND@LHC under conservative assumptions and compare them with constraints from direct detection, beam-dump, fixed-target, and collider experiments. We also assess the validity of the effective field theory description by considering conservative bounds on the couplings.
\end{abstract}

\maketitle

\pagenumbering{arabic}

\section{Introduction}

Despite the remarkable success of the Standard Model in describing particle interactions at the electroweak scale, the absence of clear signals of new physics at high energies has motivated increasing interest in scenarios involving weakly coupled dark sectors. In many well-motivated extensions of the Standard Model, new degrees of freedom interact only feebly with ordinary matter, rendering them difficult to access in conventional collider searches. Such states are instead more naturally probed in high-intensity experiments, where large fluxes and momenta can compensate for small couplings. Experiments at the Forward Physics Facility \cite{FPF1, FPF2, FPF3, FPF4, FPF5} at the LHC and fixed-target facilities therefore offer sensitivity to sub-GeV new physics that is difficult to access with conventional central detectors.\par

Among these efforts, the Scattering and Neutrino Detector at the LHC occupies a distinct position. Although the SND@LHC is designed primarily to study neutrinos produced at the ATLAS interaction point, its geometry, target composition, and location close to the beam collision axis also render it sensitive to new particles whose dominant experimental signature arises from elastic scattering. Importantly, SND operates in a regime where neutrino backgrounds are measurable and can be constrained using data from the detector itself or from simulations performed.\par

In this work we focus on sub-GeV dark states interacting with the Standard Model through electromagnetic dipole operators. In the far-forward region of the LHC, dipolar dark states can be produced efficiently through meson decays, Drell-Yan process, and other photon-mediated pair production channels. Once produced, they may scatter in the detector material, yielding observable recoil signals that compete with neutrino-induced backgrounds. This interplay between production and scattering defines a region of parameter space where SND@LHC provides sensitivity complementary to that of other facilities and does not yield leading constraints.\par

The purpose of this work is to quantify the reach of SND@LHC for such dark states in a unified framework. In Section \ref{SecII}, we introduce the dark state model considered in this study. Section \ref{SNDsec} describes the layout of SND@LHC and presents a schematic overview of the detector along with a description of the key components relevant for the detection of neutrinos and FIPs. The results of simulations of neutrino backgrounds at the SND@LHC are also presented. In Section \ref{prodsec}, we discuss the production channels relevant to our model, including meson decay widths and the resulting spectra of mesons and dark states. We compute the production cross-section both in the forward hemisphere and in the pseudorapidity range covered by SND@LHC. Finally, in Section \ref{senssec}, we derive constraints from the invisible decay widths of mesons and present sensitivity estimates for SND@LHC, comparing them with existing bounds from several experiments operating at different energy scales.

\section{Dirac Dark States and the Dipolar Model}
\label{SecII}
In our search for dark sector phenomena, fermionic dark states have been one of our candidates in many studies. An effective field theory approach to Dirac dark matter, similar to what we have addressed here, has been conducted in \cite{Dipolar_EFT} along with additional models arising from SUSY. A $U(1)_{B-L}$ extension of the Standard Model has been studied in \cite{U(1)B-L} where the dark state $\chi$ couples to the gauge boson $Z'$. Right handed neutrinos, scalar and Dirac states acting as a viable dark matter candidate has been studied in a $U(1)_X$ extension of the Standard Model in \cite{u1c}. Such models have been further studied in \cite{u1a, u1b}. In \cite{ChiralDM}, a singlet model of Dirac dark matter has been studied where the Standard Model has been extended by two chiral fermions $\psi_L$ and $\psi_R$ that form Dirac dark matter $\psi$ followed by a real scalar field $\phi$. \par
Our work focuses on a neutral dark state that carries an electric or magnetic dipole moment as described in eqn.{\eqref{singlet}. Such a state is naturally a fermion, and we treat it as a Dirac particle.  Majorana particles have vanishing electric and magnetic dipole moments which restrict our choice of states. In the context of dark matter, we know the coupling to photons should be small enough to keep dark matter $``$dark$"$. Current experimental limits reduce the viable parameter space substantially; leading to stringent bounds on the coupling across the mass range we have considered in this work. A detailed analysis on the theory of dipolar dark matter has been addressed in \cite{DDM_basic}.\par
The interactions are mediated by dimension-5 electric and magnetic dipole moment operators under the Minimal Flavor hypothesis as described in \cite{d2021freezing} in the form of a doublet and a triplet. The possible Higgs and Z boson interactions are heavily constrained and are irrelevant to our analysis. For efficient imposition of constraints and simplification of the resulting phenomenology, we consider a singlet model as described in eqn.\eqref{singlet}. This description will be utilized in deriving constraints by considering relevant processes that contribute to the production of the dark states and the detector characteristics.
\begin{equation}
    \mathcal{L} = \dfrac{c_A^{MDM}}{\Lambda}\bar{\chi} \sigma_{\mu\nu} \chi F^{\mu\nu} + i\dfrac{c_A^{EDM}}{\Lambda}\bar{\chi} \sigma_{\mu\nu} \gamma_5 \chi F^{\mu\nu}
\label{singlet}
\end{equation}
Existing direct-detection, beam-dump, and collider searches impose stringent constraints on the dipole couplings over a broad range of $\chi$ masses, typically restricting the couplings to $\mathcal{O}(10^{-6} - 10^{-3})$ depending on the value of the mass and the interaction considered. Signatures of Dirac dark matter at direct detection experiments through interactions induced by electromagnetic form factors have been analyzed through a formalism that relates the signal rate to the form factors that includes electric and magnetic dipole moments in \cite{Ibarra:2024mpq}. A possible UV completion model of dimension-5 and 6 have been discussed in the previously mentioned reference as well. Furthermore, several works that mention UV completions of dipole models can be found in \cite{UV1, UV2, UV3}. Constraints on electromagnetic dipole interactions, including direct dipole moments for Dirac fermions as well as transition dipole moments relevant for Majorana fermions, have been studied in \cite{Masso:2009mu} in relation to several nuclear recoil experiments. Furthermore, \cite{RHNdipole} studies dipole interactions among right handed neutrinos. The sensitivity of several experiments to dipole interactions using Majorana states and sterile neutrinos has been studied in \cite{dipoleexp}.

\section{The Scattering and Neutrino Detector}
\label{SNDsec}
The Scattering and Neutrino Detector(SND) is a newly designed facility at the LHC that has been designed to detect neutrinos of all three flavors and aims to take measurements of neutrinos in an untapped parameter space. Located 480 m away from the ATLAS IP in the TI18 tunnel, and positioned slightly off-axis covering the pseudorapidity range of $7.2 < \eta < 8.4$, it aims to study neutrinos originating mostly from charmed hadron decays. SND@LHC is expected to be sensitive in measuring scatterings of Feebly Interacting Particles(FIPs) \cite{fips1, fips2} in its target. A schematic diagram of the detector has been provided in Fig.\eqref{fig:2}.
\begin{figure}[!ht]
\centering
\begin{minipage}[ht]{0.707\textwidth}
\includegraphics[width = 1.0\textwidth]{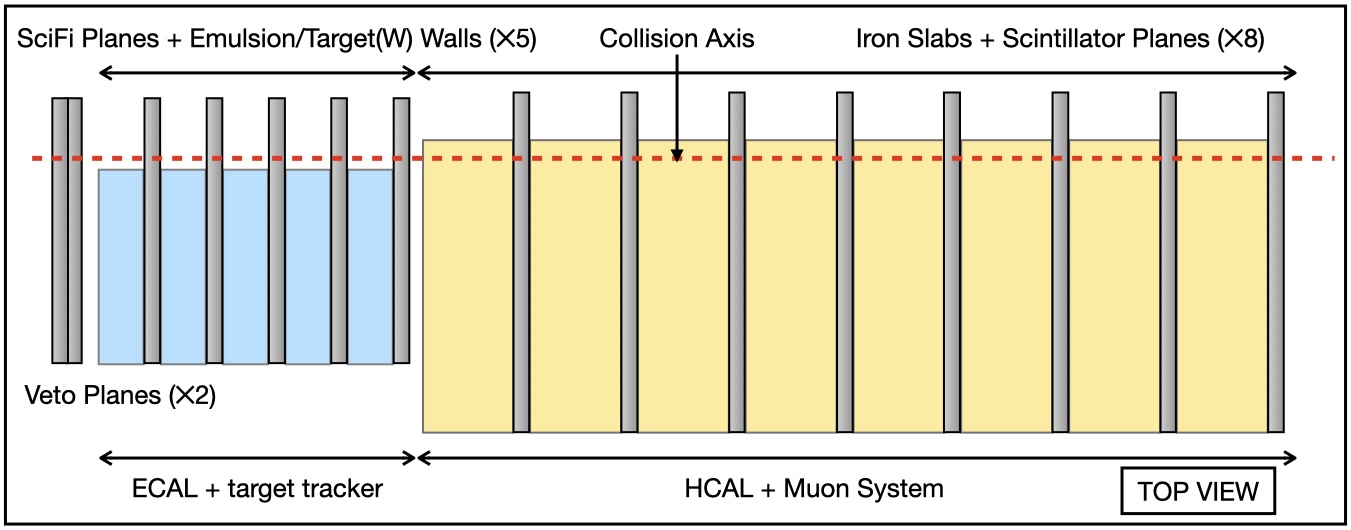}
\end{minipage}
\hfill
\begin{minipage}[ht]{0.278\textwidth}
\centering
\includegraphics[width = 1.0\textwidth]{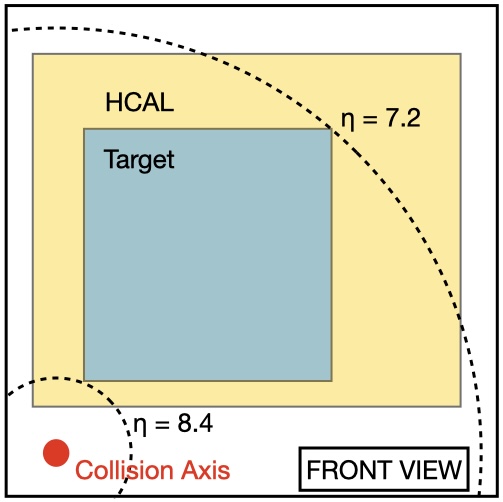}
\end{minipage}
\caption{\textbf{Left:} Schematic layout of the SND@LHC from a top view. \textbf{Right:} Frontal view of the target and HCAL with the collision axis.} 
\label{fig:2}
\end{figure}

\subsection{Structure of SND@LHC}
A detailed description of the working and assembly of the detector components can be found in \cite{SND_recent} and \cite{Ahdida:2750060}. The veto system detects charged particles by scintillation. In our analysis, we consider the target and vertex detector to register signals from the dark sector interacting with the tungsten target. The target system comprises five 40 $\times$ 40 cm$^{2}$ bricks of emulsion cloud chambers (ECC) alternated with Scintillating fiber (SciFi) plates. Each emulsion plate contains 60 emulsion films with 59 1mm thick tungsten plates alternately positioned between the films. The ECCs facilitate micrometric accuracy of any charged particle tracks and reconstruct interactions of neutrinos and other events that occur within an emulsion brick. The scintillating fiber plates are designed to measure energies of electromagnetic and hadronic showers with a spatial and temporal resolution of 50 $\mu$m and 100 ps respectively. The muon system consists of eight scintillating planes interleaved between 20 cm thick iron slabs. It will identify muon neutrinos and hence, muon neutrino charged current interactions \cite{Boyarsky:2021moj}. Combined with the SciFi planes, it will measure energies of hadronic showers.
 
\subsection{How sensitive is the SND@LHC?}
Sensitivity projections for dark sector scenarios relevant to SND@LHC have been investigated in prior works and hence, providing quantitative reach estimates in the relevant parameter spaces. The sensitivity of the SND@LHC in probing leptophobic dark matter coupled to quarks via a vector mediator has been studied in \cite{Boyarsky:2021moj}. Detection prospects of complex scalar and Majorana fermionic dark matter have been analyzed in \cite{SND_models} within the framework of far-forward detectors including SND@LHC. A model describing long lived light dark photon has been systematically studied in \cite{snd_dark} addressing detection at SND@LHC. We will derive the sensitivity curve for the SND@LHC corresponding to the interactions considered in this work, with the aim to illustrate the ability of the detector in probing different FIPs.\par

\begin{table}[!ht]
\begin{tabular}{|c|c|}
\hline
\textbf{Production Channel} & \textbf{Feynman Diagram} \\ \hline
\cellcolor{orange!10}\raisebox{6.0\height}{Drell-Yan: $q\ \bar{q}\ \rightarrow \ \gamma^* \ \rightarrow \  \chi\ \bar{\chi}$} & 
\begin{tikzpicture}
\begin{feynman}

\vertex (a) at (0, 1.5) {\(\bar{q}\)};
\vertex (b) at (0, -1.5) {\(q\)};
\vertex (c) at (1.8, 0);         
\vertex (v) at (4.0, 0);
\vertex (d) at (5.5, 1.6) {\(\chi\)};
\vertex (e) at (5.5, -1.6) {\(\bar{\chi}\)};

\diagram* [line width=1pt]{
  (a) -- [anti fermion] (c),
  (b) -- [fermion] (c),
  (c) -- [photon, edge label=\(\gamma^*\)] (v),
  (v) -- [fermion] (d),
  (v) -- [anti fermion] (e),
};

\filldraw [black] (v) circle (2pt);

\end{feynman}
\end{tikzpicture}\\ \hline
\raisebox{6.0\height}{Bremsstrahlung: $q \ q \ \rightarrow \ q \ q \ \gamma^* \ ; \ \gamma^* \ \rightarrow \ \chi \ \bar{\chi}$} &
\begin{tikzpicture}
\begin{feynman}
\vertex (a) at (-3.5, 1.4) {\(q\)};
\vertex (b) at (-3.5, -1.4) {\(q\)};
\vertex (c) at (-1, 1);
\vertex (d) at (-1, -1);\vertex (e) at (1, 1);
\vertex (f) at (2, 2);
\vertex (g) at (4, 2) {\(\chi\)};
\vertex (h) at (4, 1) {\(\bar{\chi}\)};
\vertex (o1) at (1.2, -1.4) {\(q\)};
\vertex (o3) at (3, 0) {\(q\)};

\diagram* [line width=1pt]{
  (a) -- [anti fermion] (c) -- [anti fermion,edge label=\(q\)] (e) -- [anti fermion] (o3),
  (b) -- [fermion] (d) -- [gluon, edge label=\(g\)] (c),
  (d) -- [anti fermion] (o1),
  (e) -- [photon, edge label=\(\gamma^*\)] (f),
  (f) -- [fermion] (g),
  (f) -- [anti fermion] (h),
};

\filldraw [black] (f) circle (2pt);

\end{feynman}
\end{tikzpicture}
\\ \hline
\raisebox{6.0\height}{Mono-Jet: $q\ g\ \rightarrow q \ \gamma^* \ ; \ \gamma^* \rightarrow \ \chi \ \bar{\chi}$} & 
\begin{tikzpicture}
\begin{feynman}
  \vertex (sbar_in) at (-2.5, 1.6);   
  \vertex (v_upper) at (0, 1);     
  \vertex (v_lower) at (0, -1.2);     
  \vertex (v_right) at (2, 1);      
  \vertex (chi) at (4, 1.6);      
  \vertex (chibar) at (4, -0.2);   
  \vertex (s_out) at (2.5, -1.8);     
  \vertex (g_in) at (-2.5, -1.8);     
  
  \diagram* [line width=1pt]{
    (sbar_in) -- [fermion, edge label=$q$] (v_upper),
    (v_upper) -- [fermion, edge label=$q$] (v_lower),  
    (v_lower) -- [gluon, edge label=$g$] (g_in),
    (v_lower) -- [fermion, edge label=$q$] (s_out),
    (v_upper) -- [photon, edge label=$\gamma^*$] (v_right),  
    (v_right) -- [fermion, edge label=$\chi$] (chi),
    (v_right) -- [anti fermion, edge label=$\bar{\chi}$] (chibar),
  };
    
  \filldraw (v_right) circle (2pt);
\end{feynman}
\end{tikzpicture}
\\ \hline
\cellcolor{orange!10} Pseudoscalar Meson Decay: $P \to \gamma^* \gamma \ ;\ \gamma^* \to \chi\bar{\chi}$ & 
\begin{minipage}{0.45\textwidth}
\centering
\begin{tikzpicture}
\begin{feynman}[large]
  \vertex (a) at (0,0) {$P$};
  \vertex (b) at (2,0);
  \vertex (c) at (4,1) {$\gamma$};
  \vertex (d) at (4,-1);
  \vertex (e) at (6,-2) {$\chi$};
  \vertex (f) at (6,0) {$\bar{\chi}$};
  
  \diagram* {
    (a) -- [scalar, black, line width=1pt] (b),
    (b) -- [photon, black, edge label=\(\gamma^*\), line width=1pt] (d),
    (b) -- [photon, black, line width=1pt] (c),
    (d) -- [fermion, black, line width=1pt] (e),
    (d) -- [anti fermion, black, line width=1pt] (f)
  };
\end{feynman}
\filldraw[black] (4,-1) circle (2.5pt);
\end{tikzpicture}
\end{minipage}
\\ \hline
\cellcolor{orange!10} Vector Meson Decay: $V \to \gamma^* \to \chi\bar{\chi}$ & 
\begin{minipage}{0.45\textwidth}
\centering
\begin{tikzpicture}
\begin{feynman}[large]
  \vertex (a) at (-1,1) {$V$};
  \vertex (b) at (1,1);
  \vertex (c) at (3,1);
  \vertex (d) at (5,0) {$\chi$};
  \vertex (e) at (5,2) {$\bar{\chi}$};
  
  \diagram* {
    (a) -- [scalar, black, line width=1pt] (b),
    (b) -- [photon, black, edge label=\(\gamma^*\), line width=1pt] (c),
    (c) -- [fermion, black, line width=1pt] (d),
    (c) -- [anti fermion, black, line width=1pt] (e)
  };
\end{feynman}
\filldraw[black] (3,1) circle (2.5pt);
\end{tikzpicture}
\end{minipage}
\\ \hline
\end{tabular}
\caption{Production channels of $\chi$ and $\bar\chi$. Significant and considered channels have been shown in shaded cells. For bremsstrahlung and mono-jet processes, one of several diagrams has been shown. The two processes are sub-leading in comparison to the Drell-Yan and numerically insignificant.}
\label{channels}
\end{table}
\section{Production Channels of Dark Sector}
\label{prodsec}
All relevant direct production channels of dark state($\chi$) as shown in first three rows of Table\eqref{channels}, have been simulated on \texttt{MadGraph5\_aMC@NLO} \cite{alwall2014automated} after the model was implemented through \texttt{FeynRules} \cite{Alloul_2014}. Meson decays, as shown in bottom two rows of Table\eqref{channels}, have been analyzed using \texttt{FORESEE} \cite{Kling:2021fwx}. The projections have been computed for integrated luminosities of 250 fb$^{-1}$ and 3000 fb$^{-1}$ corresponding to SND@LHC collections during the Run-3 and Run-4 of the LHC. All production channels have been listed in Table\eqref{channels}.
\subsection{Drell-Yan process}
For masses beyond 1 GeV, we have used \texttt{MadGraph5\_aMC@NLO} \cite{alwall2014automated} to find the cross-sections of Drell-Yan process and used \texttt{MadAnalysis 5} \cite{Conte:2012fm} to find the number of dark states propagating towards the SND@LHC target. Contributions from mono-jet and bremsstrahlung are substantially weaker and are safe to ignore. The obtained results presented in Fig.\eqref{Direct_prod}, will be later used in the following sections to obtain the projection of the detector.
\begin{figure}[!ht]
    \centering
    \begin{minipage}[b]{0.49\textwidth}
        \centering
        \includegraphics[width=\linewidth]{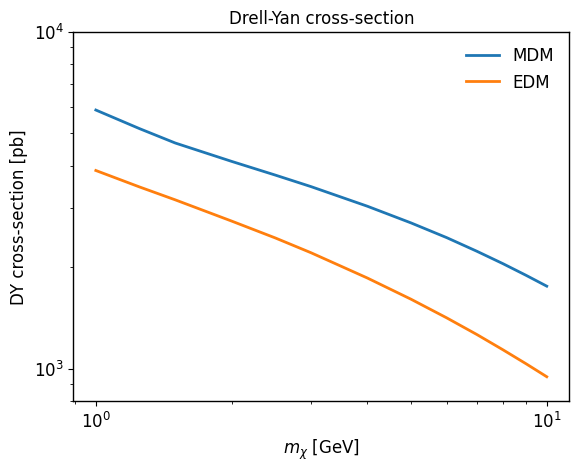}
    \end{minipage}
    \hfill
    \begin{minipage}[b]{0.49\textwidth}
        \centering
        \includegraphics[width=\linewidth]{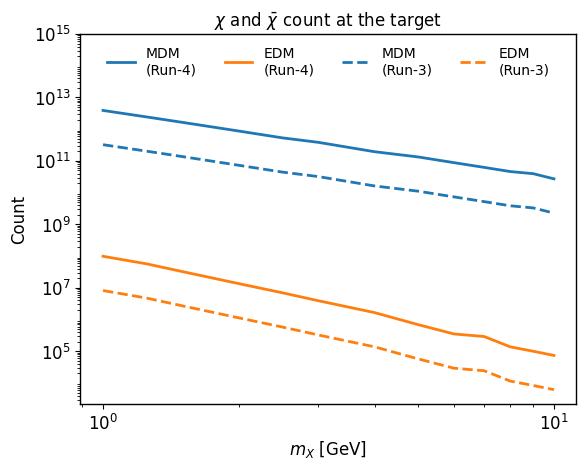}
    \end{minipage}
    \caption{\textbf{Left:} Drell-Yan production cross-section for coupling set to 1 GeV$^{-1}$. \textbf{Right:} Incident $\chi$ and $\bar\chi$ at SND@LHC target for coupling set to 1 GeV$^{-1}$ due to Drell-Yan process. The count has been normalized to both the Run-3 (250 fb$^{-1}$) and Run-4 (3000 fb$^{-1}$) data accumulation.}
    \label{Direct_prod}
\end{figure}
\subsection{Decay of Unflavored Mesons}
In the lighter-mass sector, specifically below 1 GeV, we resort to meson decays as the source of dark state production. Heavier mesons such as $\Upsilon(nS)$ and $J/\Psi$ allow us to dive into a segment of the mass region($\lesssim$ 5 GeV) of the Drell-Yan process. In this subsection, we investigate meson decays and impose constraints based on the invisible width of the mesons. We assume the entire invisible width is available for decays into dark state pairs.
\enlargethispage{\baselineskip}
\subsubsection{Meson Decay Widths}
We calculate the decay widths and branching ratios of mesons into $\chi\bar{\chi}$ pairs according to the model \eqref{singlet} as mentioned in \cite{Chu:2020ysb} with individual treatment of MDM and EDM components. We then proceed to constrain the resulting values with the invisible branching fractions as presented in \cite{ParticleDataGroup:2024cfk}.\par
The branching ratio of the vector mesons($V$) into $\chi\bar{\chi}$ pairs can be computed from,
\begin{equation}
    BR(V\rightarrow\chi\bar{\chi}) = BR(V\rightarrow e^-e^+)\dfrac{f_\chi(s_{\chi\bar\chi}^2)}{f_e(m_V^2)}\sqrt{\dfrac{m_V^2-4m_\chi^2}{m_V^2-4m_e^2}}
    \label{vxx}
\end{equation}\par
The decay width of pseudoscalar mesons ($P$) can be calculated from,
\begin{equation}
    \Gamma_\chi = \int_{4m_\chi^2}^{m_P^2} ds_{\chi\bar{\chi}}\Gamma_{P\rightarrow \gamma \gamma^*(s_{\chi\bar{\chi}})} \dfrac{f_\chi(s_{\chi\bar{\chi}}^2)}{16\pi^2s_{\chi\bar{\chi}}^2}\sqrt{1-\dfrac{4m_\chi^2}{s_{\chi\bar{\chi}}}}
    \label{pxx}
\end{equation}
where $s_{\chi\bar\chi}$ is the invariant mass squared of the $\chi\bar\chi$ pair and $\Gamma_{P\rightarrow \gamma \gamma^*}$ is the decay width of the pseudoscalar meson into a photon pair; with one photon being off-shell.
\begin{equation}
    \Gamma_{P\rightarrow \gamma \gamma^*} = \dfrac{\alpha^2 (m_P^2-s_{\chi\bar{\chi}})^3}{32\pi^3m_P^3F_P^2}
\end{equation}
where $F_P$ is the decay constant of the meson and can be sourced from \cite{Colquhoun:2015oha},\cite{Chang:2018aut} and \cite{PhysRevD.54.1}. The functions $f_\chi$ and $f_e$ are stated below:
\begin{equation}
    MDM: f_\chi(s_{\chi\bar{\chi}}) = \dfrac{8}{3}(c_A^{MDM})^2s_{\chi\bar{\chi}}^2\left(1+\dfrac{8m_\chi^2}{s_{\chi\bar{\chi}}}\right)
    \label{fxmdm}
\end{equation}

\begin{equation}
    EDM: f_\chi(s_{\chi\bar{\chi}}) = \dfrac{8}{3}(c_A^{EDM})^2s_{\chi\bar{\chi}}^2\left(1-\dfrac{4m_\chi^2}{s_{\chi\bar{\chi}}}\right)
    \label{fxedm}
\end{equation}

\begin{equation}
    f_e(m_V^2)=\dfrac{16\pi\alpha}{3} m_V^2 \left(1+\dfrac{2m_e^2}{m_V^2}\right)
    \label{fe}
\end{equation}\par 

\subsubsection{Cross-Sections and Spectra}
We employ \texttt{FORESEE} \cite{Kling:2021fwx} to simulate the production of dark state $\chi$ from the decays of unflavored vector and pseudoscalar mesons. \texttt{FORESEE} is a package that enables us to simulate different production and decay modes of BSM states when pre-defined interactions along with geometric cuts specific to a detector are applied by the user. It also includes spectra of several mesons generated through the appropriate event generators in the forward region. \texttt{FORESEE} also allows us to simulate over different collision energies and have a choice of generator for certain mesons.\par
We have shown representative examples of meson spectra obtained in Fig. \eqref{MesSpec} using \texttt{FORESEE} pre-generated spectra. The spectra of lighter mesons($\omega, \rho, \phi, \eta, \eta', \pi^0$) have been obtained through the \texttt{EPOS LHC} \cite{EPOSLHC} generator implemented in the \texttt{CRMC} \cite{CRMC} interface, while heavier mesons ($J/\psi, \Psi(2S), \Upsilon(nS)$) have been generated through \texttt{Pythia 8} \cite{bierlich2022comprehensive} and tuned to LHCb \cite{LHCbjpsi, LHCb2s, LHCbups} data. The forward meson spectra used in this analysis are subject to generator-dependent systematic uncertainties. These uncertainties affect the predicted dark state flux and hence, alter the resulting sensitivity rather than qualitatively modifying its characteristic properties. 
\begin{figure}[!ht]
    \centering
    \begin{minipage}[ht]{0.50\textwidth}
        \centering
        \includegraphics[width=\linewidth]{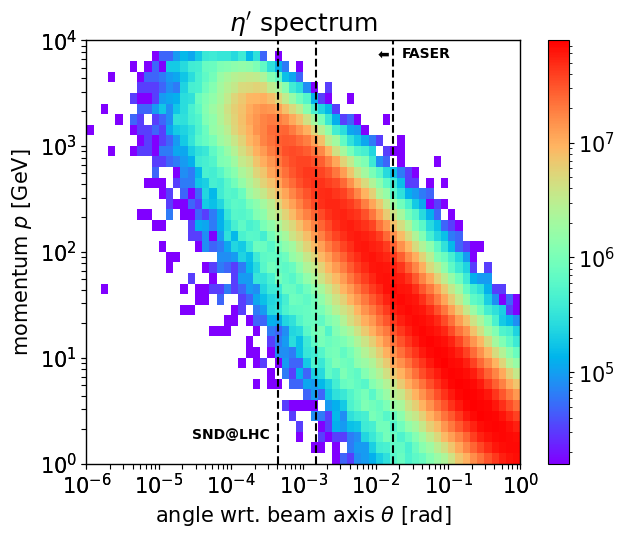}
    \end{minipage}
    \hfill
    \begin{minipage}[ht]{0.49\textwidth}
        \centering
        \includegraphics[width=\linewidth]{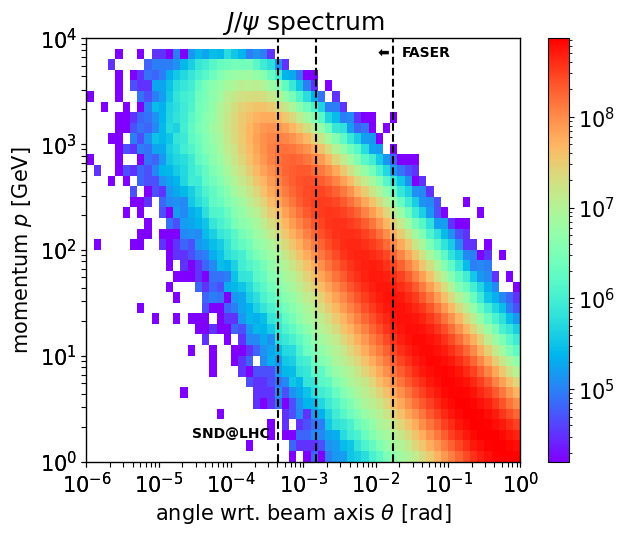}
    \end{minipage}
    \caption{\textbf{Left:} $\eta'$ spectrum using \texttt{EPOS LHC} generator. \textbf{Right:} $J/\psi$ spectrum using \texttt{Pythia 8} generator. The angular coverage of SND@LHC and FASER \cite{faser_1, faser_2} in the spectra has been shown with black dotted lines for comparison. The color map represents the cross-section in pb per bin on a logarithmic scale.}
    \label{MesSpec}
\end{figure}\par
Our following study has been conducted by treating the electric and dipole moments individually. Eqn.\eqref{vxx} has been used to input the branching ratio of the vector mesons. We can apply the differential branching ratios \cite{Flare} of pseudoscalar mesons to simulate the 3-body decays into $\gamma \chi \bar{\chi}$.

\begin{align}
    &\frac{\mathrm{d}BR^{(MDM)}_{P \rightarrow \gamma \chi \bar{\chi}}}
         {\mathrm{d}q^2\,\mathrm{d}(\cos\theta)} 
    = BR_{P \rightarrow \gamma \gamma} \left[\frac{(c_A^{MDM})^2}{4\pi^2} \left(1-\frac{q^2}{M^2}\right)\sqrt{1 - \frac{4m_\chi^2}{q^2}}\left(\frac{8m_\chi^2}{q^2} + \left[1 - \frac{4m_\chi^2}{q^2}\right]\sin^2\theta\right)\right]\\
    &\frac{\mathrm{d}BR^{(EDM)}_{P \rightarrow \gamma \chi \bar{\chi}}}
         {\mathrm{d}q^2\,\mathrm{d}\cos\theta} 
    = BR_{P \rightarrow \gamma \gamma} \left[\frac{(c_A^{EDM})^2}{4\pi^2} \left(1-\frac{q^2}{M^2}\right)\sqrt{1 - \frac{4m_\chi^2}{q^2}}\left(1 - \frac{4m_\chi^2}{q^2}\sin^2\theta\right)\right]
\end{align}

where, $q$ is the squared invariant mass of the $\chi\bar\chi$ pair, $M$ is the mass of the decaying pseudoscalar meson, $\theta$ is the 
angle between the direction of $p_\chi$ in $\chi\bar\chi$-pair rest frame and $p_\chi+p_{\bar\chi}$ in the rest frame of the meson.\par
In the analysis that follows, we have fixed our coupling to 1 GeV$^{-1}$. The $\chi\bar{\chi}$ pair production cross-section from decays of mesons has been obtained through a p-p collision at 14 TeV for unit coupling and shown in Fig.\eqref{mes_cs}. 
\begin{figure}[!ht]
    \centering
    \begin{minipage}[b]{0.49\textwidth}
        \centering
        \includegraphics[width=\linewidth]{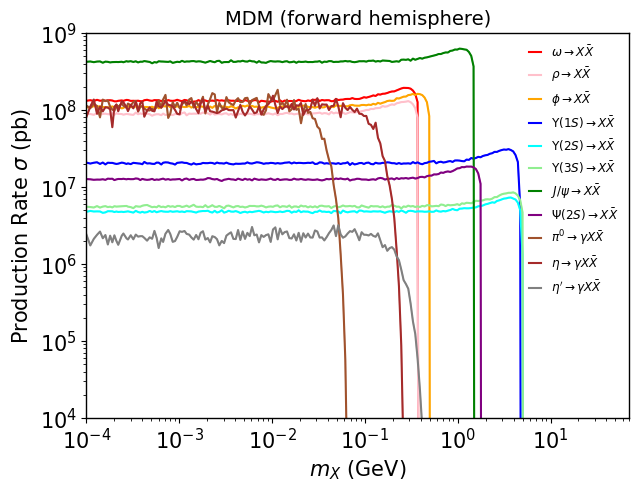}
    \end{minipage}
    \hfill
    \begin{minipage}[b]{0.49\textwidth}
        \centering
        \includegraphics[width=\linewidth]{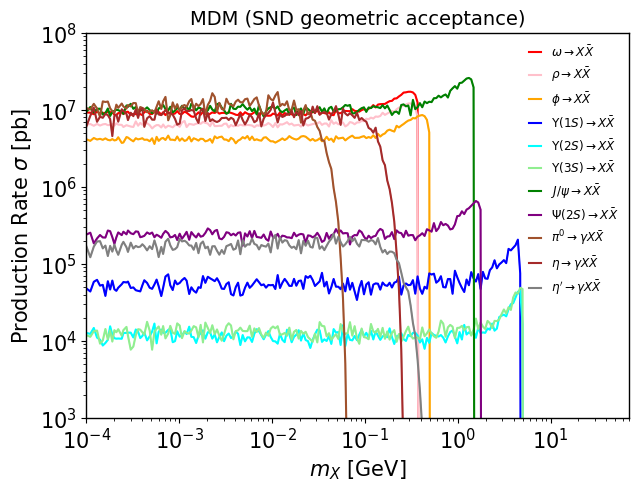}
    \end{minipage}
    \begin{minipage}[b]{0.49\textwidth}
        \centering
        \includegraphics[width=\linewidth]{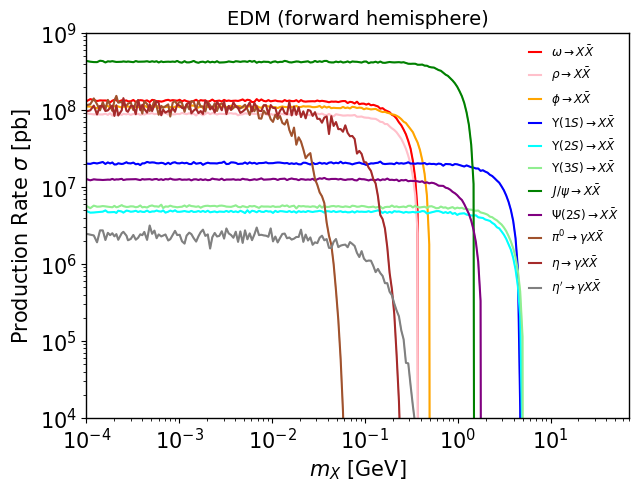}
    \end{minipage}
    \hfill
    \begin{minipage}[b]{0.49\textwidth}
        \centering
        \includegraphics[width=\linewidth]{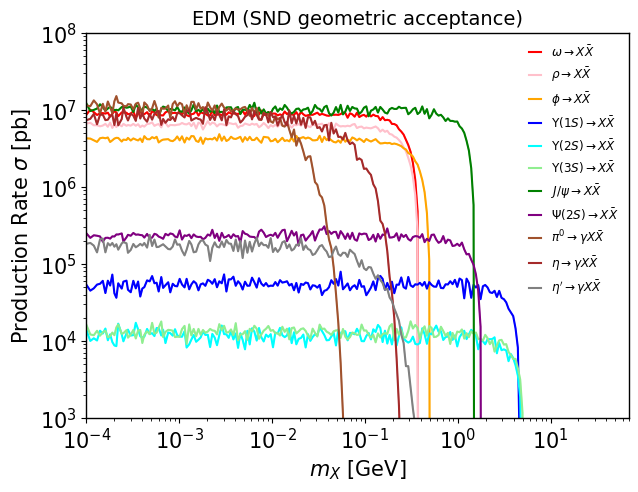}
    \end{minipage}
    \caption{We present here the $\chi\bar\chi$ pair production cross-section both in the half hemisphere and the annular region described by the pseudorapidity coverage of SND@LHC. \textbf{Top Left and Top Right:} MDM cross-section in forward hemisphere and within 7.2 < $\eta$ < 8.4. \textbf{Bottom Left and Bottom Right:} EDM cross-section in forward hemisphere and within 7.2 < $\eta$ < 8.4.}
    \label{mes_cs}
\end{figure}
In our analysis, we have considered the unflavored mesons available in \texttt{FORESEE}. In the cross-sections computed above, we have treated magnetic and electric dipole moments separately. MDM and EDM cross-sections behave similarly in low mass limit and start differing near threshold masses. A similar model-independent analysis has been conducted in Appendix \ref{app_modind} for similar decay channels. As expected, the cross-sections of production of $\chi\bar\chi$ pairs from meson decays approach zero when $m_\chi$ equals half the mass of the decaying meson. The ratio of combined incident $\chi$ and $\bar\chi$ in the region spanned by $\eta$ and $\phi$ cuts for SND@LHC to the area within 7.2 < $\eta$ < 8.4 is 0.176. The $\chi e$ scattering events, which has been used to analyze the sensitivity of the SND@LHC to the dipole model, can be calculated as following as a function of energy $E$ of the incoming flux of FIPs:
\begin{equation}
    N_{events} = \sigma_{prod}(E)\; \mathcal{L}\; Z\; l_{det}\; \sigma_{scatt}(E)
\end{equation}
where $\sigma_{prod}$ are production cross-section from meson decays and Drell-Yan process, $\sigma_{scatt}$ is the relativistic scattering cross-section at the target of the detector, $\mathcal{L}$ is the integrated luminosity, $Z$ is the atomic number of the target material, which is tungsten for SND@LHC, $l_{det}$ is the length of the target material available for scattering.\par 
A momentum spectrum has been obtained for dark states after having defined the dynamics of our model in Fig.\eqref{LLP_spec}. In direct detection experiments, the dark state spectrum is significant in determining the recoil energy spectrum at the target of the detector. The DM spectra can help distinguish different models, which might be searched for at the facility. In addition, the spectra can help us optimize the structure of the detectors that would be used to probe these models.
\begin{figure}[!ht]
    \centering
    \begin{minipage}[b]{0.49\textwidth}
        \centering
        \includegraphics[width=\linewidth]{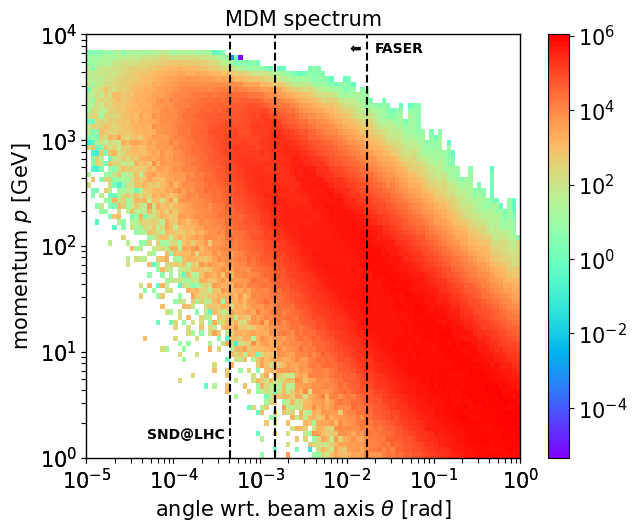}
    \end{minipage}
    \hfill
    \begin{minipage}[b]{0.49\textwidth}
        \centering
        \includegraphics[width=\linewidth]{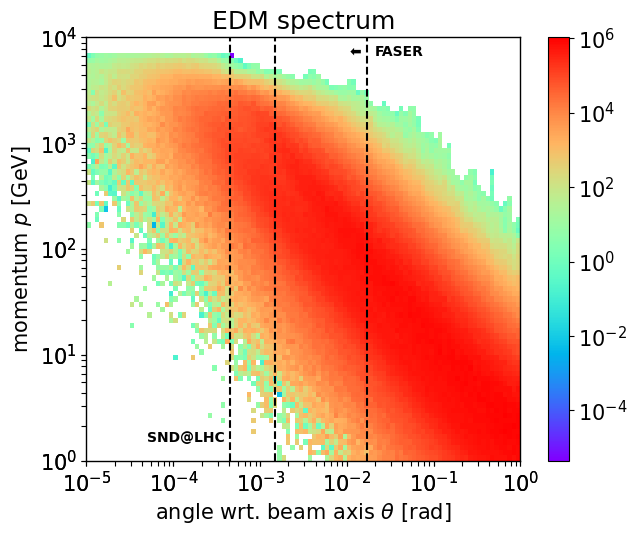}
    \end{minipage}
    \caption{Spectrum for $m_\chi$ = 0.1 GeV with coupling set to 1 GeV$^{-1}$. \textbf{Left:} MDM; \textbf{Right:} EDM. The angular coverage of SND@LHC and FASER in the spectra has been shown with black dotted lines for comparison. The color map represents the cross-section in pb per bin on a logarithmic scale.}
    \label{LLP_spec}
\end{figure}

A distinct spectrum corresponding to different production channels has also been shown in Fig.\eqref{momentaspec} for a Monte Carlo sampling of 100 per point in the meson spectra. Such momentum spectra tell us the expected recoil energies at the target when considering specific resonance channels. The approximate similarity of spectra obtained in Fig.\eqref{momentaspec} can be traced back to Fig.\eqref{LLP_spec} that shows minimal difference as well.
\begin{figure}[!htbp]
    \centering
    \begin{minipage}[ht]{0.49\textwidth}
        \centering
        \includegraphics[width=\linewidth]{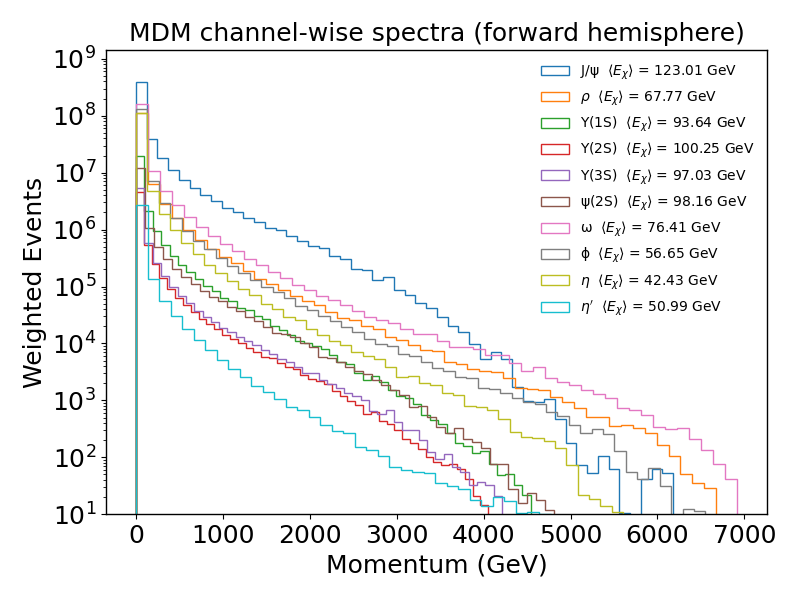}
    \end{minipage}
    \hfill
    \begin{minipage}[ht]{0.485\textwidth}
        \centering
        \includegraphics[width=\linewidth]{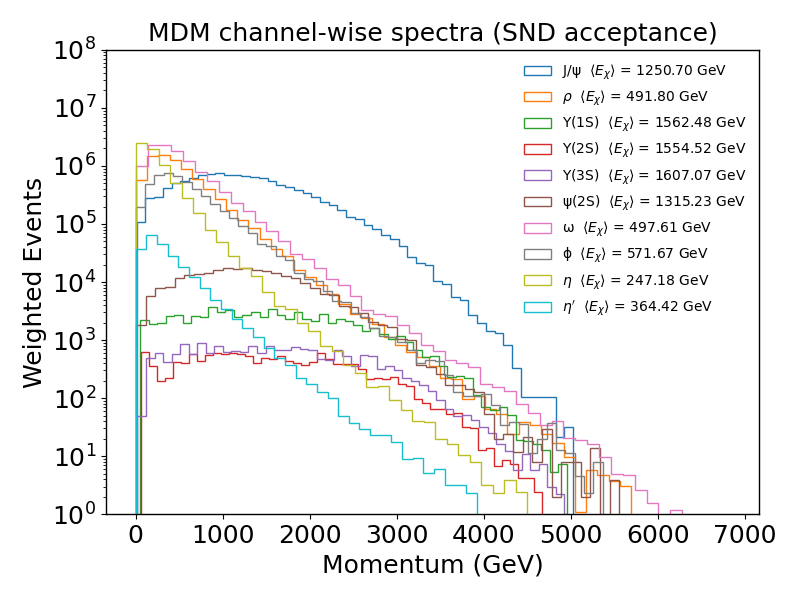}
    \end{minipage}
    \centering
    \begin{minipage}[ht]{0.49\textwidth}
        \centering
        \includegraphics[width=\linewidth]{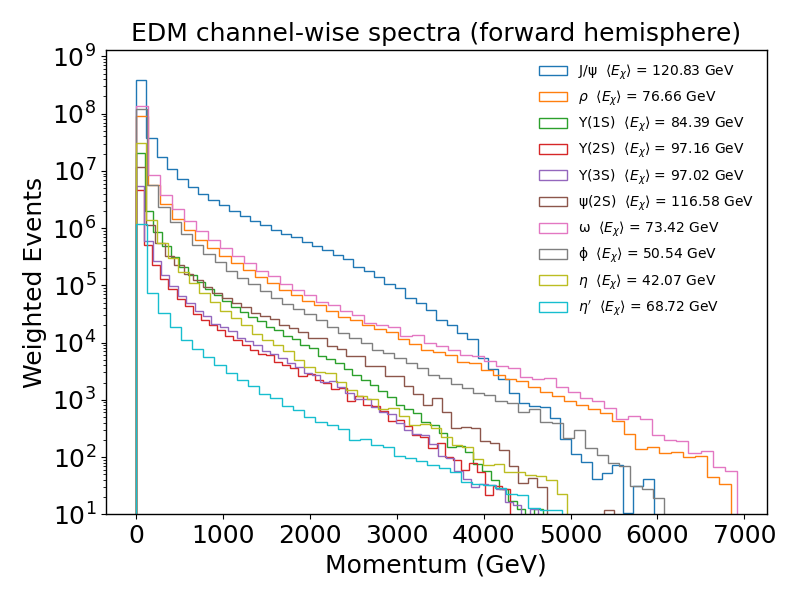}
    \end{minipage}
    \hfill
    \begin{minipage}[ht]{0.485\textwidth}
        \centering
        \includegraphics[width=\linewidth]{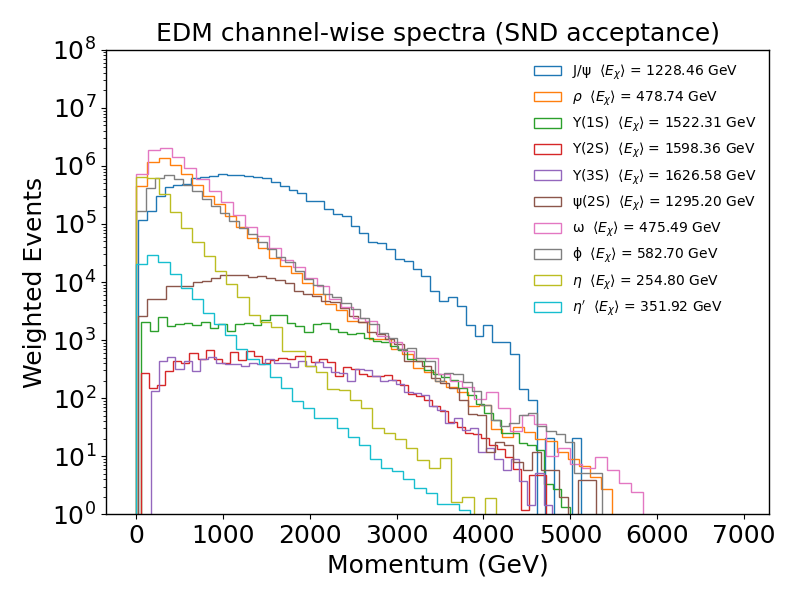}
    \end{minipage}
    \caption{Weighted events (cross-section) as a function of momentum for $m_\chi$ = 0.1 GeV for different channels of production. a) MDM spectrum in the forward hemisphere. b) MDM spectrum within 7.2 < $\eta$ < 8.4. c) EDM spectrum in the forward hemisphere. d) EDM spectrum within 7.2 < $\eta$ < 8.4. Along with the histograms we also mention the average momenta of $\chi$ produced from each meson decay for both situations. The data have been acquired with the coupling set to 1 GeV$^{-1}$.}
    \label{momentaspec}
\end{figure}

\section{Neutrino Background at SND@LHC}
The scattering of FIPs off protons is mimicked by neutral current (NC) neutrino scattering events. The background is partly contributed by inelastic scattering of neutrinos on protons through neutral-current(NC) scattering. In \cite{Ahdida:2750060}, the neutrino background was simulated using the offline software framework \texttt{sndsw}, that was developed by the SHiP \cite{SHiP} collaboration by utilizing the \texttt{FairRoot} \cite{fairroot} framework. \texttt{sndsw} integrates several high-energy physics software packages that allow a detailed simulation of detector signals and backgrounds. Across all flavors, the number of NC scattering events is found to be 450 upon normalization to an integrate luminosity of 150 fb$^{-1}$. In \cite{SND_recent}, the neutrino interactions have been simulated using \texttt{Genie} \cite{Genie} which results in 550 NC interactions at the SND@LHC target for an integrated luminosity of 250 fb$^{-1}$. Charged current interactions have also been discussed in above referred works but are irrelevant to our study.\par
For elastic scattering of FIPs off nucleons, the background is provided by neutrino NC resonant and deep inelastic scattering events where the number of single track events is expected to be 1.7 \cite{Boyarsky:2021moj} during the entire time of operation. However, the background from reaction products that cannot be associated with the muon or neutrino vertex has not been studied yet. In the case of elastic scattering off electrons, the background largely comes from neutrinos scattering on electrons and can be ignored safely once visibility cuts on corresponding tracks produced in neutrino scattering are taken into consideration \cite{Ahdida:2750060}. Neutrinos scattering on electrons has been widely studied in \cite{nue_2001, nue_2003, nue_2012, nue_2020}. The contributions of both these elastic processes to the background can be further reduced by a selective analysis of topology and kinematics as demonstrated in \cite{top_kin}.\par
Neutrinos scattered in the rock placed in front of the SND@LHC produce pions and leptons. Pions can be separated from the electrons scattered by the FIP with the help of the micrometric accuracy of the emulsion cloud chambers. Muons are recognized as the most penetrating particle by the electromagnetic calorimeter and identified by the muon system located downstream. Tau decays can be identified topologically by their decay vertex. However, other sub-dominant contributions from neutrons and $K_L$'s, can be studied further and have not been considered here.\par
In this study, we do not consider charged current (CC) contribution to backgrounds from neutrino-electron scattering assuming the vertex reconstruction works with unit efficiency. A neutrino interacting via CC interaction with a nucleon, produces a lepton which can mimic an electromagnetic shower due to scattering with an FIP; however, an efficient identification of this neutrino-nucleon vertex can omit this source of background.\par
SND@LHC can utilize its time of resolution which is of $\sim$200ps to perform model-independent searches for FIPs by combining a recoil signature with a time of flight measurement to suppress the background comprising neutrino interactions. It is expected that SND@LHC will be able to resolve scattering signals of massive FIPs from neutrino scattering events in certain regions of a parameter space spanned by the momentum and mass of the FIP as discussed in \cite{Ahdida:2750060}. However, the average energy of FIPs corresponding to our model is too large to belong within the region of the parameter space in which such a search for FIP can be performed. Hence, detection by time of flight measurement is not applicable in the detection of our model described in eqn.\eqref{singlet}.

\section{Sensitivity of SND@LHC}
\label{senssec}
The sensitivity of SND@LHC in the detection of scalar dark matter has already been investigated in \cite{Boyarsky:2021moj}. It also discusses the sensitivity of the detector in probing decays of dark scalars, dark photons and heavy neutral leptons. We compute projected limits on the detection of our current model containing a potential dark matter via scattering on electrons at the target assuming near-unit detection efficiency, motivated by the strong electromagnetic shower reconstruction ability. \par
An efficient way to impose a limit on the coupling comes from studying the invisible decay widths of mesons which can be found in \cite{PDG2024}. The condition that our model has to produce $\chi\bar{\chi}$ pairs within the allowed invisible decay widths of each meson, allows us to put an upper bound on the allowed couplings as illustrated in Fig.\eqref{Mes_inv}. We consider decays of mesons for which we have an experimental record of their invisible branching ratios. The bounds placed by $\Upsilon(1S)$ are by far the most stringent. Future improvements in the experimental measurements of invisible branching ratios will affect our limits.\par
\begin{figure}[!htbp]
    \centering
    \begin{minipage}[b]{0.49\textwidth}
        \centering
        \includegraphics[width=\linewidth]{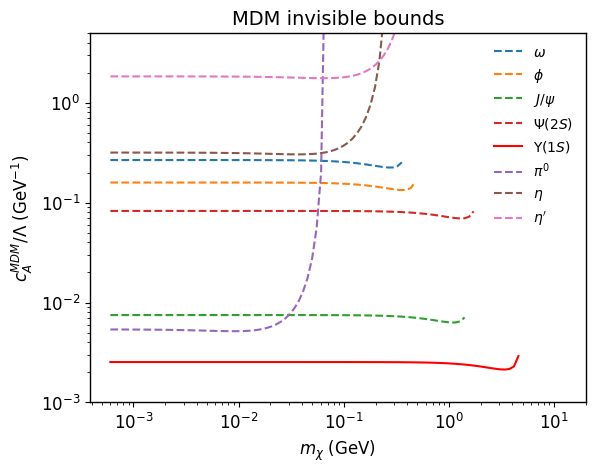}
    \end{minipage}
    \hfill
    \begin{minipage}[b]{0.49\textwidth}
        \centering
        \includegraphics[width=\linewidth]{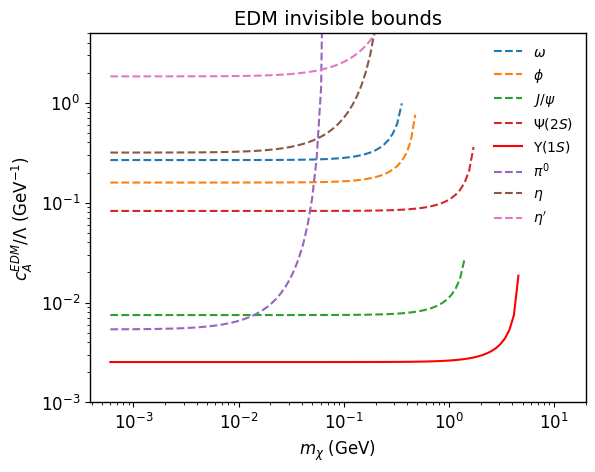}
    \end{minipage}
    \caption{Bounds derived from inv. B.R. of decaying mesons. \textbf{Left:} MDM. \textbf{Right:} EDM}
    \label{Mes_inv}
\end{figure}\par
Ref. \cite{LDMemfactors} also addresses bounds placed by rare meson decays such as $K^+ \rightarrow \pi^+ \chi \bar\chi$ and $B^+ \rightarrow K^+ \chi \bar\chi$. The $K^+$ decay constraints $c_A^{MDM}$ or $c_A^{EDM} \lesssim 1.5 \times 10^{-4}\mu_B$ in certain momentum ranges when the mass of $\chi$ is in the MeV scale while the $B^+$ decay puts the constraint $c_A^{MDM}$ or $c_A^{EDM} \lesssim 3 \times 10^{-3}\mu_B$, where $\mu_B$ is the Bohr magneton.\par
An essential part of obtaining the sensitivity projection of a detector includes the energies and the population of the $\chi$ or $\bar{\chi}$ population incident on the target. We take into account both direct production and meson decays as sources of $\chi/\bar{\chi}$ production and project the average energies and population for unit coupling as functions of $m_\chi$ as shown in Fig.\eqref{en_num}. The variation of $\langle E_\chi \rangle$ can be related to the production cross-section from different channels and their spectra for a particular mass. The energy plot shows dips at points where the mass exceeds the kinematic limits of specific mesons. For example, at $m_\chi\sim$ 1.5 GeV, the $J/\psi$ channel becomes kinematically forbidden and the energy sees a dip because $J/\psi$ exhibits higher weighted events compared to other mesons over a significant region of the momentum-region. Although Drell-Yan process allows the produced $\chi\bar\chi$ pairs to attain higher energies, the plot sees a dip at $m_\chi\sim$ 4.7 GeV for the kinematic limits corresponding to $\Upsilon(nS)$ mesons are crossed. The energy keeps increasing beyond that point because Drell-Yan produces heavier particles with larger momenta.\par
In the dark state population plot, the contribution from meson decays and Drell-Yan has been added. We see a dip at $m_\chi\sim$ 1.5 GeV because the decay channel of $J/\psi$ becomes inaccessible kinematically. The contribution from meson decays drops to zero after $m_\chi\sim$ 4.7 GeV when the $\Upsilon(nS)$ channels become kinematically forbidden. The overlap of Drell-Yan contribution masks the clear dips in meson decay contribution. Beyond $m_\chi \approx$ 4.7 GeV, Drell-Yan dominates the curve for both interactions. Overall, the population curve inherits the characteristics from the cross-section plots in Fig. \eqref{mes_cs}.
\begin{figure}[!htbp]
    \centering
    \begin{minipage}[b]{0.49\textwidth}
        \centering
        \includegraphics[width=\linewidth]{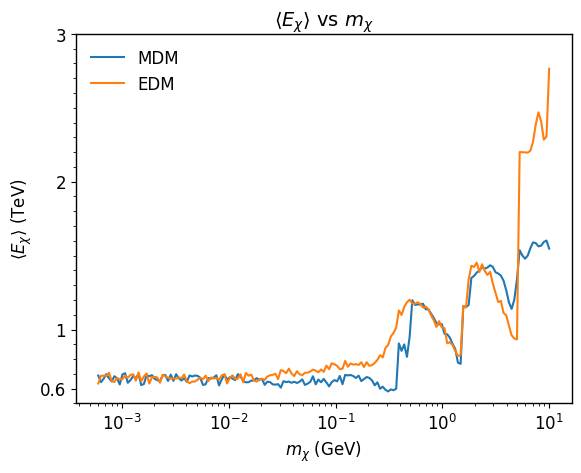}
    \end{minipage}
    \hfill
    \begin{minipage}[b]{0.50\textwidth}
        \centering
        \includegraphics[width=\linewidth]{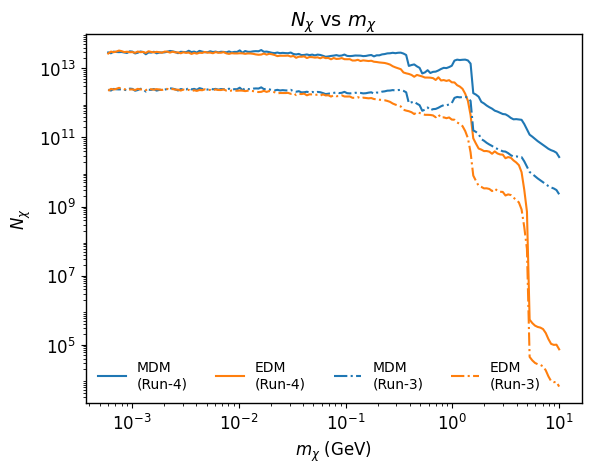}
    \end{minipage}
    \caption{\textbf{Left:} Average energy $\langle E_\chi \rangle$ in SND@LHC acceptance varied with mass. The average has been computed over all channels for each specific value of $m_\chi$. \textbf{Right:} Incident $\chi$ and $\bar\chi$ populations at the target of SND@LHC from the ATLAS IP. The populations have been normalized to both the Run-3 (250 fb$^{-1}$) and Run-4 (3000 fb$^{-1}$) data accumulations. In both plots, we have considered all relevant channels of production.}
    \label{en_num}
\end{figure}
\subsection{$\chi N$ scattering}
Several experiments probing DM-nucleus scattering have been conducted producing limits on the corresponding cross-section. Direct detection experiments looking into DM-nucleus cross-sections in the mass range of few GeV to several TeV have significantly constrained the parameter space. Experiments like LUX \cite{LUX}, PandaX-II \cite{PandaX-N}, XENON1T \cite{XENON1T-N}, DarkSide-50 \cite{DarkSide-N}, CDMSlite \cite{CDMS-N}, CDEX-0 \cite{CDEX0-N}, CDEX-1 \cite{CDEX1a-N, CDEX1b-N, CDEX1c-N, CDEX1d-N}, CDEX-10 \cite{CDEX10a-N, CDEX10b-N}, CDEX-1B \cite{CDEX1Ba-N, CDEX1Bb-N} probe the GeV scale and beyond. In the sub-GeV region, which is of interest to us, experiments like CRESST-III, XQC, CDEX-1B and CDEX-10 \cite{subGeVa, subGeVb, subGeVc} have placed constraints. However, such experiments enforce limits on the per-nucleon cross-section which is translated to DM-nucleus cross-section under the assumption that the interaction mediating the scattering is momentum-independent. This underlying condition of drafting the cross-section limits forbids us from deriving constraints on both MDM and EDM interactions.\par
When considering scattering on nuclei, we need to take into account the momentum dependence of both MDM and EDM interactions. The different dependence on recoil energies has been explicitly stated in Appendix \ref{app_relcs}. This necessitates the presence of MDM and EDM scattering cross-section limits from other experiments in order to draw a comparison of $\chi$N sensitivity of SND@LHC with other experiments. The $\chi N$ elastic and deep inelastic scattering cross-sections for both interactions with $c/\Lambda = 1$ GeV$^{-1}$ are $\sim$ 9.48$\times$10$^{-31}$ cm$^2$ and 4$\times$10$^{-29}$ cm$^2$ respectively for a minimum recoil energy of 100 MeV and can be used as benchmark points for comparison with other $\chi N$ scattering experiments.

\subsection{$\chi e$ scattering}
Several direct detection experiments have probed elastic scattering on electrons in a wide mass region from 10s of keV to several GeV. Constraints from these experiments such as these allow us to compare the prospects of our model at SND@LHC. Direct detection experiments with semiconductor targets \cite{SENSEI, DAMIC, EDELWEISS, CDMShvev, HvevOLD}, noble liquid-based detectors \cite{XENON1T, XENON10_100, DarkSide, XENON10_old, DarkSide, PandaX} have provided limits on DM-electron elastic scattering cross-sections. Some EDM constraints corresponding to protoSENSEI@MINOS and CDMS-HVeV have been obtained from \cite{proto1, proto2}. Ref. \cite{CDEXcomp} presents a list of DM-electron cross-section limits that have been mentioned here as well.\par
Halo dark matter with a mass around 10 MeV has a kinetic energy of about 5 eV which can be amplified by scattering off electrons in the Sun and being reflected towards the Earth with an energy in the keV range \cite{SRDM1}. This allows us to probe lighter dark matter more efficiently compared to halo origins. We adopt constraints derived from solar reflected dark matter(SRDM) derived from XENON1T \cite{SRDM1, SRDM2, SRDM3, Emken} in the low mass region that we are interested in.\par
DM-electron cross-sections are expressed in terms of a reference cross-section defined as \cite{refcs}:
\begin{equation}
    \bar\sigma_e = \frac{1}{16\pi(m_e + m_\chi)^2}\overline{|\mathcal{M}_{\chi e}|}^2_{q^2 = \alpha^2m_e^2}
\end{equation}
where,
\begin{equation}
    \overline{|\mathcal{M}_{\chi e}(q)|}^2 = \overline{|\mathcal{M}_{\chi e}(q_0)|}^2_{q_0^2 = \alpha^2m_e^2} \times |F_{DM}(q)|^2
\end{equation}\par
We are considering the squared amplitudes for scattering of DM off free electrons after averaging over initial states and summing over final states. The leading terms of the amplitudes $|\mathcal{M}_{\chi e}|^2$ are laid down below:
\begin{align}
    &\text{MDM:} \quad
    64\pi \frac{{c_A^{MDM}}^2}{\Lambda^2}\alpha m_\chi^2\\
    &\text{EDM:} \quad
    \frac{256\pi {c_A^{EDM}}^2\alpha m_e^2 m_\chi^2}{q^2\Lambda^2}
\end{align}\par
Hence, $F_{DM} = 1$ for MDM and $\alpha m_e/q$ for EDM. We explore the limits on $\bar\sigma_e$ that feature the same form factors. Experimental limits on $\chi e$ cross-section have been established by several direct detection experiments and have been shown in Fig.\eqref{DD_limits}.
\begin{figure}[!ht]
    \centering
    \begin{minipage}[b]{0.50\textwidth}
        \centering
        \includegraphics[width=\linewidth]{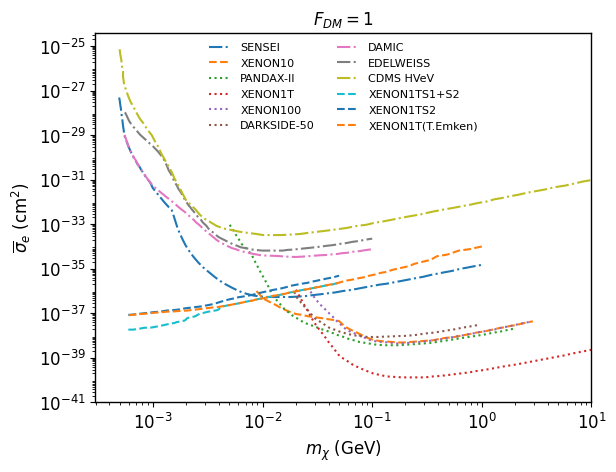}
    \end{minipage}
    \hfill
    \begin{minipage}[b]{0.49\textwidth}
        \centering
        \includegraphics[width=\linewidth]{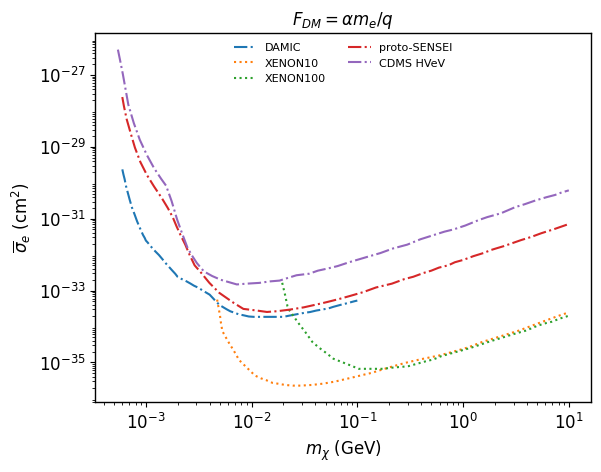}
    \end{minipage}
    \caption{Upper limits on $\chi e$ reference cross-section $\bar\sigma_e$ corresponding to 90$\%$ C.L. \textbf{Left:} Upper limits on cross-sections corresponding to form factor $F_{DM}$ = 1. \textbf{Right:} Upper limits on cross-sections corresponding to form factor $F_{DM}$ = $\alpha m_e / q$.}
    \label{DD_limits}
\end{figure}

Applying the projected dynamics of $\chi$ shown at the start of this section, we have produced a sensitivity curve for SND@LHC that can be put against current direct detection limits as shown in Fig.\eqref{snd_DD}. Assuming a background-free scenario and no observed events, the 90$\%$ C.L. upper limit on the expected number of signal events is obtained from Poisson statistics as 2.3 events. We adopt this criterion to define the projected sensitivity, corresponding to the minimum signal yield required to reach 90$\%$ C.L. In the projected limits, we observe that the curve strongly depends on the DM population for the variation in $\chi e$ cross-section is subdominant compared to flux variations. We assume that the minimum detectable energy of the electron recoils is 100 MeV. 
\begin{figure}[!ht]
    \centering
    \begin{minipage}[b]{0.49\textwidth}
        \centering
        \includegraphics[width=\linewidth]{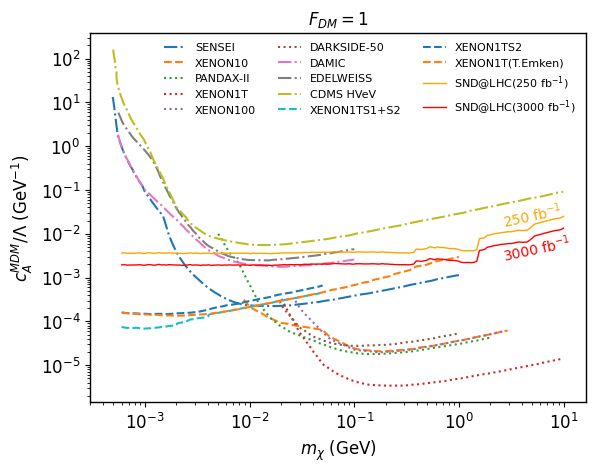}
    \end{minipage}
    \hfill
    \begin{minipage}[b]{0.49\textwidth}
        \centering
        \includegraphics[width=\linewidth]{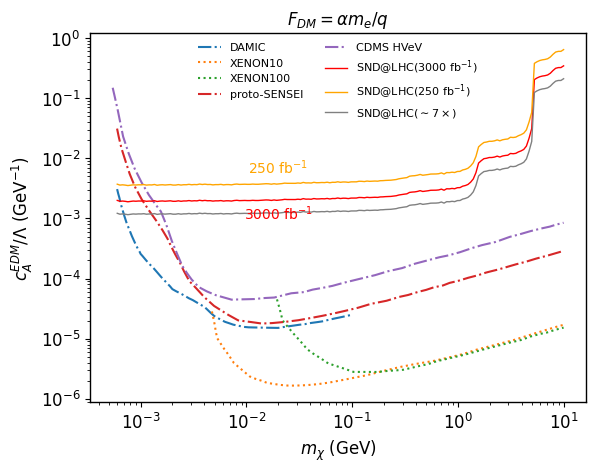}
    \end{minipage}
    \caption{Projected sensitivity plot of SND@LHC shown against: (Left) 90$\%$ C.L. constraints on MDM couplings fixed by DD experiments ($F_{DM}$=1). (Right) 90$\%$ C.L. constraints on EDM couplings fixed by DD experiments ($F_{DM}=\alpha m_e/q$). Sensitivity reaches of SND@LHC corresponding to both the Run-3 (250 fb$^{-1}$) and Run-4 (3000 fb$^{-1}$) have been shown. The gray line corresponds to an approximate $\sim$7$\times$ enhancement in EDM yield, as discussed in the next section.}
    \label{snd_DD}
\end{figure}\par

The constraints from direct detection experiments are derived at energies $\sim$ $\mathcal{O}$(keV), while SND@LHC operates at several hundreds of GeV to a few TeV. In order to obtain a fair comparison, we evolve the low energy constraints via the appropriate RGE equations as shown in Appendix \ref{app_rgerunning}. However, the change in constraints is insignificant when considering the RGE evolution of dimension-5 MDM and EDM operators. MDM and EDM, being CP-even and CP-odd respectively, do not mix with each other, and the evolution is entirely governed by the running of SM parameters.\par
We present the sensitivity plot along with constraints provided by beam-dump, fixed-target, and electron-positron collider experiments separately in Fig.\eqref{snd_beam}. Experiments such as LEP \cite{LDMemfactors}, BaBar \cite{BaBar}, CHARM II \cite{CharmII}, E613 \cite{E613}, LSND \cite{LSND}, and MiniBooNE \cite{MiniBooNE1, MiniBooNE2} have imposed bounds on both MDM and EDM bounds. A compilation of several such bounds has also been presented in \cite{BeamComp}. The corresponding projections at future electron colliders have been studied in \cite{future_electron_colliders}. Since such experiments operate at much higher scale of transfer momentum, there is no requirement of applying the renormalization group equations for a valid comparison with projections at SND@LHC. Although there are indirect and cosmological constraints on the dipole moments as discussed in \cite{LDMemfactors, indirect1, indirect2}, we do not mention them in Fig.\eqref{snd_DD} and Fig.\eqref{snd_beam}. Apart from experimental bounds, \cite{LDMemfactors, Flare} also looks into the generated relic abundance for both MDM and EDM interactions. Future upgraded versions \cite{AdvSND, AdvSND1, SND@HL-LHC} of SND@LHC are expected to have improved projections aimed at different dark sectors, including the Dirac dipole model.
\begin{figure}[!htbp]
    \centering
    \begin{minipage}[ht]{0.49\textwidth}
        \centering
        \includegraphics[width=\linewidth]{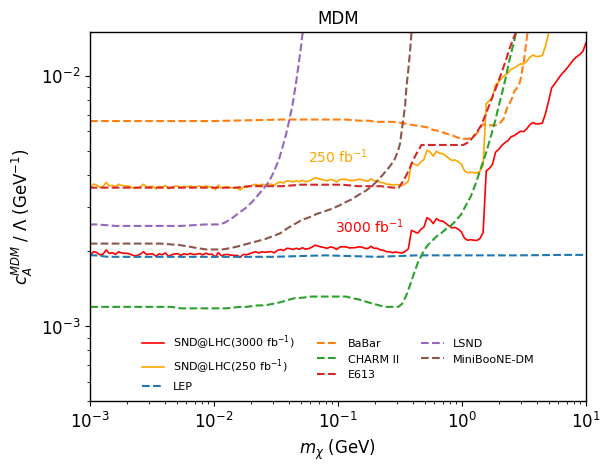}
    \end{minipage}
    \hfill
    \begin{minipage}[ht]{0.49\textwidth}
        \centering
        \includegraphics[width=\linewidth]{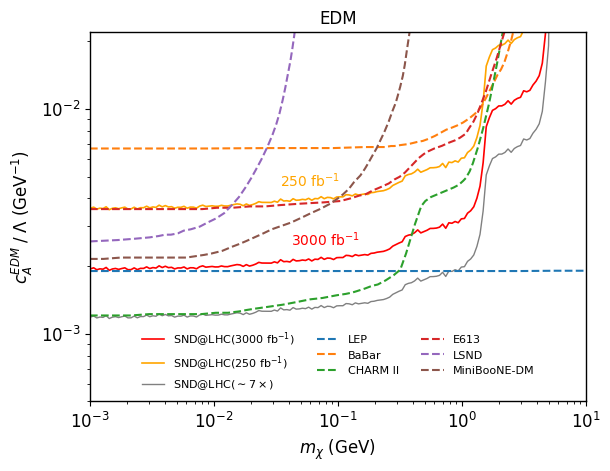}
    \end{minipage}
    \caption{Projected sensitivity plot of SND@LHC shown against constraints from beam-dump, fixed-target and $e^-e^+$ collider experiments. Left: 90$\%$ C.L. constraints on MDM couplings. Right: 90$\%$ C.L. constraints on EDM couplings. Sensitivity reaches of SND@LHC corresponding to both the Run-3 (250 fb$^{-1}$) and Run-4 (3000 fb$^{-1}$) have been shown. The gray line corresponds to an approximate $\sim$7$\times$ enhancement in EDM yield, as discussed in the next section.}
    \label{snd_beam}
\end{figure}\par
We proceed to present the region of EFT validity for both MDM and EDM interactions in Fig.\ref{eftval}. To assess the regime of validity of the effective description, we introduce a representative momentum transfer $q_{rep}$, defined from the kinematics of $\chi e$ scattering using the average energy of the incoming $\chi$ for a given mass $m_\chi$. We impose a conservative condition $\frac{c\; q_{rep}}{\Lambda} < \frac{1}{10}$ to ensure that the effective interactions are perturbative, where c is the Wilson coefficient assumed to be $\mathcal{O}(1)$ and $\Lambda$ is the UV scale of the EFT. We emphasize that this should be interpreted as a consistency criterion based on a representative momentum scale rather than a strict bound derived from the maximum kinematically accessible momentum transfer.
\begin{figure}[!htbp]
\centering
\includegraphics[width = 0.50\linewidth]{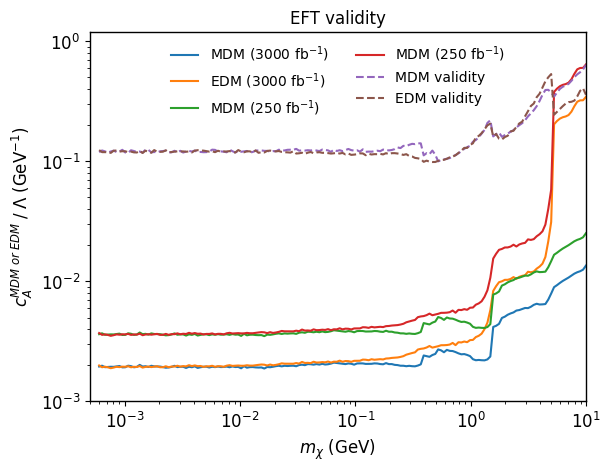}
\caption{EFT validity region of MDM and EDM sensitivities. In this figure, we have considered the projected reach corresponding to both 250 fb $^{-1}$ and 3000 fb$^{-1}$.}
\label{eftval}
\end{figure}

\section{Conclusions and Outlook}

We have analyzed the prospects for detecting fermionic dark states at SND@LHC in the benchmark scenario where the leading interaction with the Standard Model is described by electric and/or magnetic dipole operators. Propagating the resulting fluxes to
the SND@LHC target, we evaluated the scattering signal rates for nuclei and electrons and presented projected sensitivities for LHC integrated luminosities relevant to both the current and HL-LHC program.\par
Our results show that SND@LHC can probe dipole moments in a manner complementary to existing experimental bounds comprising direct-detection, beam-dump, and collider searches, for sub-GeV to few-GeV dark matter, while potential future enhancements in incident flux or detector target area can further extend the accessible parameter space. We have also studied the parameter space for which the effective dipole description remains perturbative at the momentum transfers relevant for scattering at the target region. We identify the domains where the EFT remains applicable even with a strict validity condition imposed.\par
As seen from the comparison between the SND@LHC sensitivity and the existing direct detection constraints, solar reflected dark matter severely restricts the low-mass region of the MDM parameter space. Probing beyond the currently excluded region would require an enhancement of nearly four orders of magnitude in the effective LLP exposure over the HL-LHC scenario, allowing SND@LHC to begin probing the MDM parameter space around $m_\chi\sim10^{-2}$ GeV. However, SND@LHC in the HL-LHC era can probe the low-mass region for EDM more effectively than DAMIC around $m_\chi \sim$ 0.5 MeV. Extending the sensitivity of SND@LHC to even lower masses or increasing the effective LLP event yield allows access to regions of parameter space currently unconstrained by present direct detection experiments.\par
A comparison of the SND@LHC with beam-dump, fixed-target and $e^-e^+$ collider searches reveals that LEP and CHARM II currently provide the strongest limits to the Dirac dipole model. To overcome the bounds imposed by the LEP and/or CHARM II on the MDM parameter space, increasing the LLP event yield to twice the HL-LHC expectation would allow SND@LHC to begin probing previously unconstrained regions around $m_\chi\sim1-2$ GeV. Furthermore, amplifying the event yield by approximately a factor of 3.5 relative to the HL-LHC estimate would extend the sensitivity of SND@LHC in the EDM parameter space to unconstrained regions around $m_\chi\sim0.3$ GeV. Such improvements could be achieved by widening the target region or increasing the interaction length within the detector. In order to access previously unexplored regions of EDM parameter space, after taking into account all existing bounds, including direct detection limits, we must enhance the yield by a factor just exceeding 7. However, the volume available in the TI18 tunnel upstream of the detector provides scope for modifications to the current detector configuration.\par
The present analysis has been performed for the currently installed SND@LHC design. In order to achieve the aforementioned improvements to the event yield, one may exploit the available space upstream of the detector. An inspection of the layout of the region occupied by the detector indicates that a fraction of the upstream tunnel region is occupied by the electronics racks and the cooling infrastructure, rather than active detector elements. However, these components are transportable installations and could in principle be relocated within the available area inside the tunnel if needed.\par

Consequently, the following layout-driven modifications to the detector can allow an increase in the length of the target material available for scattering:} 
\begin{itemize}
\item Upon adding additional emulsion bricks and SciFi planes, the ECAL length could be increased several-fold, potentially by $\sim$ 2.4 m. This may increase the expected signal yield by a factor of 7 compared to the original configuration ($\sim$ 0.4 m \cite{SND40cm}). For reference, we may consider the AdvSND \cite{AdvSND, advconf1, advconf2}, where the detector is proposed to be placed in an excavation extending 5.7 m horizontally and 6.7 m along the length of the tunnel.
\item Furthermore, excavations of the tunnel floor in the region of the detector placement could permit a vertical extension of the target region from 40 cm to more than 1 m, thereby improving the geometric acceptance by a factor of approximately 3 and allowing a larger incident FIP flux at the target. AdvSND \cite{AdvSND, advconf1, advconf2} has suggested similar modifications, wherein the detector is to be placed in an excavation that is up to 1.25 m deep, while SND@HL-LHC {\cite{SND@HL-LHC}} proposes an excavation much shallower than that considered for AdvSND to improve the physics reach. However, an excavation requires rerouting of the drainage system.
\item Either of these possibilities or a combination thereof, could provide a viable route towards enhancing the sensitivity of the detector. Such modifications would naturally require a corresponding extension of the neutron shield and/or thermal enclosure to ensure stable operation of the ECAL.
\end{itemize}
The proposed ECAL extension of SND@LHC in this work would increase the total length of the detector to $\sim$5 m, while the AdvSND occupies a length of $\sim$6.2 m and the SND@HL-LHC \cite{SND@HL-LHC}, a length of $\sim$3 m. In incorporating the above modifications, one should take into account the changes in the pseudorapidity coverage of the target region, which may significantly affect the observed FIP spectra.\par

It is worth noting that the discussed upgrades to the detector only enable probing the EDM parameter space around $m_\chi <$ 670 keV. In contrast, the MDM parameter space remains severely constrained by the SRDM limits. In the absence of these constraints, the above mentioned enhancements would have enabled SND@LHC to probe unconstrained regions of the MDM parameter space.\par
Since the event count is proportional to the target length, an increase in the scattering region could significantly improve the sensitivity to the dark states discussed here. Such possibilities provide further motivation for deeper exploration of BSM searches at the SND@LHC.

\begin{acknowledgments}
The author expresses his gratitude to Sudhir Kumar Vempati and Biplob
Bhattacherjee for discussions and guidance and also for going through the manuscript. The author thanks Felix Kling and Sebastian Trojanowski for their discussion on \texttt{FORESEE} and electromagnetic form factors. The author would also like to thank Aadarsh Singh for his initial assistance with software. The author acknowledges financial support from Ministry of Human Resource Development, Government of India, via Prime Minister Research Fellowship (PMRF).
\end{acknowledgments}
\appendix

\section{Relativistic Elastic Cross-Section}
\label{app_relcs}
In a relativistic treatment, the recoil differential elastic cross-section for a nucleus with charge Z, electric and magnetic form factors $F_E$ and $F_M$, mass $m_N$
and spin $I_N$ when scattering on a fermion $\chi$ with mass $m_\chi$
and energy $E_\chi$, assumes the form \cite{LDMemfactors}:
\begin{equation}
    \dfrac{\mathrm{d}\sigma}{\mathrm{d}E_R} =\frac{1}{A}\Big[g_E(E_R)\alpha Z^2F_E^2(t) + g_M(E_R)\dfrac{\mu_N^2m_N^2}{\pi}\frac{I_N + 1}{3I_N}F_M^2(t)\Big]
    \label{XNelastic}
\end{equation}
where $A = (E_\chi^2 - m_\chi^2)(2m_N+E_R)$ and $t = 2m_NE_R$. The functions $g_E(E_R)$ and $g_M(E_R)$ are model-specific and given by:
\begin{equation}
    \begin{aligned}
    \text{MDM:}\quad
    &\frac{g_E}{4(c_A^{MDM})^2} = \frac{m_N}{2E_R}(4E_\chi^2-4m_\chi^2+E_R^2) - (m_\chi^2+2m_NE_\chi)\\
    &\frac{g_M}{4(c_A^{MDM})^2} = \frac{E_R}{2m_N}(m_\chi^2-m_N^2-2m_NE_\chi)+(E_\chi^2+m_\chi^2)
    \end{aligned}
\end{equation}
\begin{equation}
    \begin{aligned}
    \text{EDM:}\quad
    &\frac{g_E}{4(c_A^{EDM})^2} = \frac{m_N}{2E_R}(E_R-2E_\chi)^2\\
    &\frac{g_M}{4(c_A^{EDM})^2} = -\frac{E_R}{2m_N}(m_\chi^2+m_N^2+2m_NE_\chi)+(E_\chi^2-m_\chi^2)
    \end{aligned}
\end{equation}\par
In case of scattering off nuclei with charge Z, the electric form factor is given by,
\begin{equation}
    F_E(t)=\frac{Za^2(Z)t}{(1+a^2(Z)t)(1+t/d(A))}
\end{equation}\par
where, $A$ is the mass number, $a(Z)=\frac{111Z^{1/3}}{m_e}$ and $d(A)=0.164A^{-2/3}$. For heavy targets like tungsten, we neglect the magnetic form factor. In case of scattering off free electrons, we set $F_E = F_M = 1$, $Z=1$, $I_N = \frac{1}{2}$ and $\mu_N=\mu_B$. $m_i$ being the mass of the scattered target particle, the upper limit of integration in eqn.\eqref{XNelastic} is given by,
\begin{equation}
    E_R^{max} = \frac{2m_i(E_\chi^2-m_\chi^2)}{m_i(2E_\chi+m_i)+m_\chi^2}
\end{equation}\par

\section{Renormalization Group Running}
\label{app_rgerunning}
The Wilson coefficients are derived at very low energies $\sim \mathcal{O}$(keV) and have to be evolved to a higher energy scale $\sim$ few hundred GeV to 1 TeV. The renormalization evolution of the Wilson coefficients is governed by the anomalous dimension matrix $\gamma$ as described in \cite{RGE}:
\begin{equation}
    \frac{\mathrm{d}\vec{C}(\mu)}{\ln\mu} = \gamma^T\vec{C}(\mu)
    \label{rge}
\end{equation}\par
where, $\vec{C}(\mu)$ is the column of Wilson coefficients appearing in the renormalization group equation. However, MDM and EDM operators, being CP-even and CP-odd respectively, do not mix into each other and evolve independently. In order to keep the anomalous dimension matrices free of gauge couplings, we rescale our Wilson coefficients by $\alpha_1/2\pi$, where $\alpha_1=g_1^2/4\pi$, $g_1$ being the $U(1)_Y$ coupling. Without the presence of any other CP-even or CP-odd operators, the evolutions of $c_A^{MDM}$ and $c_A^{EDM}$ are dictated by only the $\gamma(1,1)$ element. We will treat the anomalous dimension matrices as corresponding to an electroweak singlet. Since the evolution of both $c_A^{MDM}$ and $c_A^{EDM}$ are given by the same matrix, we will denote the coefficients as just $c$. Proceeding from eqn.\eqref{rge}, with the above mentioned criteria, we arrive at,
\begin{equation}
    \frac{1}{2}\frac{\mathrm{d}(\alpha_1c)}{\mathrm{d}\ln\mu} = \frac{1}{2}\frac{\alpha_1}{4\pi}\frac{41}{3}\alpha_1c
    \label{rge1}
\end{equation}\par
The evolution of the $U(1)_Y$ coupling is given by,
\begin{equation}
    \frac{\mathrm{d}g_1}{\mathrm{d}\ln\mu}=\frac{1}{16\pi^2}\frac{41}{6}g_1^3
    \label{gevol}
\end{equation}\par
Substituting eqn.\eqref{gevol} in eqn.\eqref{rge1}, we conclude that $c$ does not evolve in the absence of other operators with same CP characteristics. Hence,
\begin{equation}
    \frac{\mathrm{d}c}{\mathrm{d}\ln\mu} = 0
    \label{cevol}
\end{equation}\par
The operator basis mentioned in \cite{RGE} can be broken down to extract the interaction with the photon as follows:
\begin{equation}
\begin{aligned}
    \mathcal{Q}_1^{(5)} &= \frac{g_1}{8\pi^2}(\bar\chi \sigma_{\mu\nu}\chi)B_{\mu\nu}\\
    &=\frac{e}{8\pi^2}(\bar\chi \sigma_{\mu\nu}\chi)F_{\mu\nu} - \frac{g_1s_w}{8\pi^2}(\bar\chi \sigma_{\mu\nu}\chi)Z_{\mu\nu}
\end{aligned}
\end{equation}\par
where, $e$ is the electric charge given by $g_1c_w$; $c_w$ and $s_w$ are the cosine and sine of the Weinberg angle respectively. Since we ignore the Z-channel in our analysis due to its negligibly available invisible width, we choose to work only with the photon vertex. According to eqn.\eqref{cevol}, $c$ does not evolve; hence, the evolution of this vertex is entirely governed by the SM-evolution of $e$ which is given as follows \cite{erunning}: 
\begin{equation}
    e^2=\frac{\hat{e}^2(M_Z)}{1+(\hat{\alpha}/\alpha)\Delta\hat{\alpha}(M_Z^2)}
\end{equation}\par
where, $\hat{e}$ and $\hat{\alpha}$ are the $\overline{MS}$ electric charge and the QED coupling constant respectively. $\Delta\hat{\alpha}(M_Z^2)$ is a function expressed in terms of photon polarization functions. After calculation of 2-loop electroweak contribution to the Thomson scattering amplitude, the values $\alpha^{-1}$($M_Z$) has been found at different scales as shown in Table\eqref{RGE}.
\begin{table}[ht]
\centering
\begin{tabular}{|l|l|}
\hline
$\mu$(GeV) & $\alpha^{-1}$($\mu$) \\ \hline
91.187 & 128.122 \\ \hline
300 & 127.369 \\ \hline
500 & 127.046 \\ \hline
800 & 126.748 \\ \hline
1000 & 126.607 \\ \hline
\end{tabular}
\caption{$\alpha^{-1}$($\mu$) computed at different scales with 2-loop electroweak corrections \cite{erunning}.}
\label{RGE}
\end{table}

\section{Model-Independent Cross-Section Analysis in \texttt{FORESEE}}
\label{app_modind}
\begin{figure}[!ht]
    \centering
    \begin{minipage}[t]{0.49\textwidth}
        \centering
        \includegraphics[width=\linewidth]{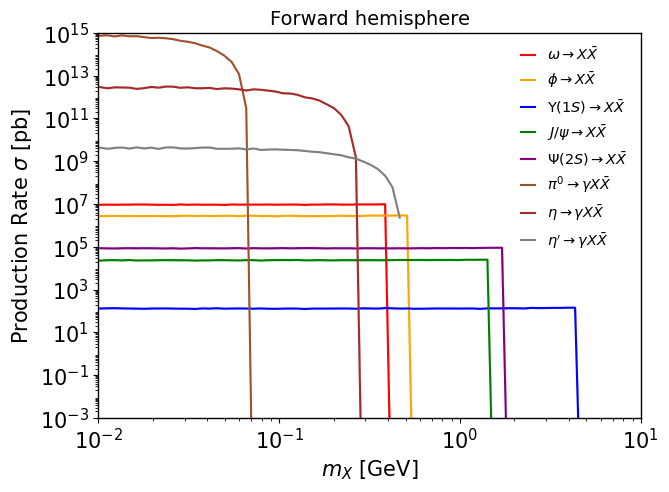}
    \end{minipage}
    \hfill
    \begin{minipage}[t]{0.49\textwidth}
        \centering
        \includegraphics[width=\linewidth]{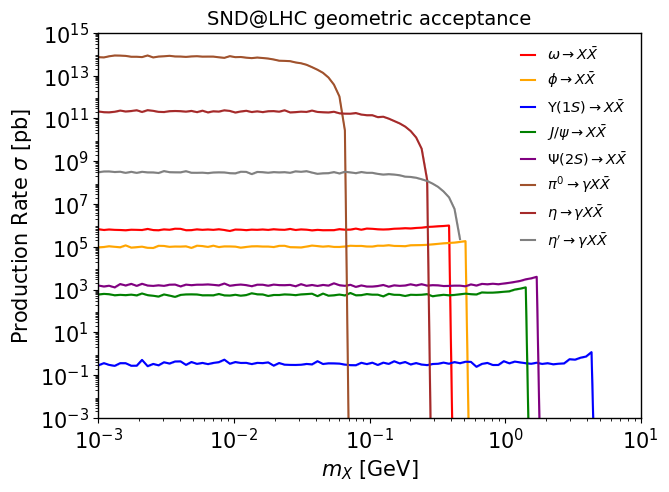}
    \end{minipage}
    \caption{Production rate of $\chi\bar\chi$-pairs in (\textbf{Left}) the forward half-hemisphere and (\textbf{Right}) in the geometric acceptance of 7.2 < $\eta$ < 8.4.}
    \label{MIcs}
\end{figure}

We present here a model-independent benchmark estimate of the upper limit of production rates of dark states originating from neutral unflavored mesons that can serve as inputs for a broad class of BSM scenarios. We equate the branching ratio of decaying mesons into $\chi\bar\chi$ pairs to their invisible branching ratio while pseudoscalar mesons need special treatment due to their three-body decays that demands the input of differential branching ratio for accurate analysis in \texttt{FORESEE}. In this benchmark analysis, we assume a constant amplitude given by $|\mathcal{M}|^2 = \epsilon^2$ (GeV$^2$), which serves as a simplified parameterization of the decay rate of pseudoscalar mesons; while for vector mesons, we simply use the invisible branching ratio to calculate the production rate of dark states. This approximation is not intended to represent a realistic dynamical model but to provide an estimate of production rates consistent with existing invisible bounds on meson decays. The differential branching ratio of pseudoscalar mesons into $\gamma \chi \bar{\chi}$ is given by:
\begin{equation}
    \frac{\mathrm{d}BR(P\rightarrow\gamma\chi\bar\chi)}{\mathrm{d}q^2\mathrm{d}\cos{\theta}} = \frac{1}{512M\pi^3}\mathcal{|M|}^2\sqrt{1-\frac{4m_\chi^2}{q^2}}\Big(1-\frac{q^2}{M^2}\Big)
\end{equation}\par
where, $M$ is the mass of the decaying pseudoscalar meson, $q$ is the squared invariant mass of the $\chi\bar\chi$ pair, $\theta$ is the 
angle between the direction of $p_\chi$ in $\chi\bar\chi$-pair rest frame and $p_\chi+p_{\bar\chi}$ in the rest frame of the meson.  $|\mathcal{M}|^2=\epsilon^2$ (GeV$^2$) takes upon an allowed range of values to keep the branching fraction of the pseudoscalar mesons within the experimental bounds on invisible decays. With the above mentioned information, we plot the production rates for this model-independent scenario with $\epsilon=1$ for $\pi^0,\eta\;\text{and}\;\eta'$ in Fig.\eqref{MIcs}.
\par
\begin{figure}[!htbp]
    \centering
    \includegraphics[width=0.5\linewidth]{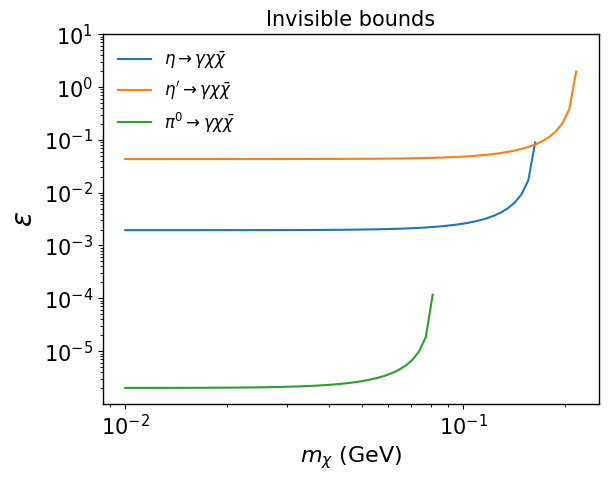}
    \caption{Model-independent invisible bounds imposed by pseudoscalar invisible branching ratios.}
    \label{ModIndbounds}
\end{figure}

As stated above, $\epsilon$ can only take values at different values of $m_\chi$ that keep the decays of pseudoscalar mesons into $\chi$ and $\bar\chi$ within their corresponding invisible branching ratios as shown in Fig.\eqref{ModIndbounds}.
\bibliographystyle{JHEP}
\bibliography{bibli.bib}
\end{document}